%% file: main.tex
\documentclass [notitlepage, floatfix, superscriptaddress, showpacs, aps,pra, 10pt]{revtex4-1}

\usepackage{amsmath}
\usepackage{hyperref}
\usepackage{graphicx}
\usepackage{bbm}
\usepackage{amssymb}
\usepackage[english]{babel}
\usepackage{lmodern}
\usepackage[T1]{fontenc}
\usepackage{wasysym}
\usepackage[normalem]{ulem}

\usepackage{dcolumn}
\usepackage{color}
\usepackage[caption=false]{subfig} 

\graphicspath{Figurepng}

\DeclareMathOperator{\e}{\displaystyle e}
\DeclareMathOperator{\de}{\displaystyle d}

\usepackage{amssymb}

\graphicspath{{Figs/}}

\begin{document}

\title{Spin dynamic response to a time dependent field}

\author{Giuseppe Bevilacqua}

\author{Valerio Biancalana} 
\email{valerio.biancalana@unisi.it}
\affiliation{Dept. of Information Engineering and Mathematics - DIISM, University of Siena -- Via Roma 56, 53100 Siena, Italy}
\author{Yordanka Dancheva}
\affiliation{Dept. of Physical Sciences, Earth and Environment - DSFTA, University of Siena -- Via Roma 56, 53100 Siena, Italy}
\affiliation {Current address: Aerospazio Tecnologie srl, Strada di Ficaiole, 53040 Rapolano Terme (SI), Italy.}
\author{Alessandro Fregosi}
\affiliation{Dept. of Physical Sciences, Earth and Environment - DSFTA, University of Siena -- Via Roma 56, 53100 Siena, Italy}
\affiliation{ CNR Istituto Nazionale di Ottica, Via Moruzzi 1, 56124 Pisa, Italy}
\author{Antonio Vigilante}
\affiliation{Dept. of Information Engineering and Mathematics - DIISM, University of Siena -- Via Roma 56, 53100 Siena, Italy}

\begin{abstract}
  The dynamic  response of a  parametric system constituted by  a spin
  precessing in a time dependent magnetic field is studied by means of
  a perturbative approach that
  unveils  unexpected features, and is then experimentally validated.   The first-order analysis
  puts in evidence different regimes: beside a tailorable 
  low-pass-filter behaviour, a
  band-pass response with interesting potential applications emerges. 
  Extending the analysis to the second perturbation order permits to study the response to generically oriented fields and to characterize several non-linear features in the  behaviour of such kind of systems.

\end{abstract}

\date{\today}
\nopagebreak
\maketitle

\input{introduction.tex}

\input{model.tex}

\input{setup.tex}

\input{discussion.tex}

\input{conclusion.tex}


\bibliographystyle{ieeetr}

\typeout{}

\bibliography{bibliograph}

\end{document}

%% file: introduction.tex
\section{Introduction}
\label{sec:introduction}


Optical  atomic  magnetometers  (OAMs)  can be  used  to  detect  time
dependent  fields (TDF)  that can  be fast  varying, disadvantageoulsy
oriented and not necessarily small.
Several kinds  of optical  magnetometers reach their  best performance
when measuring weak  and quasi-static fields. These  systems are often
modelled and analyzed  under these conditions, and  their behaviour in
the  above mentioned,  more  general case  is  scarcely discussed  and
investigated.

The possibility of measuring magnetic fields using optical pumping and
probing of alkali atoms was pointed out
more than 60 years ago, in the  
works   of   Dehmelt   \cite{demhelt_pr_57}   and   Bell   and   Bloom
\cite{bellandbloom_pr_57,  bell_prl_61}.   In  the  late  Sixties  and
Seventies, important  further steps  were carried out  by the
group  of  Cohen-Tannoudji \cite{dupont_rpa_70,  dupont_pla_69}.   The
mentioned  works  demonstrated  the   potential  for  high  resolution
magnetometry based  on magneto-optical effects which  were known since
the   historical    observations   described   by    Michael   Faraday
\cite{faraday_1848}  in non-resonant  high density  materials, and  by
Macaluso and Corbino \cite{macaluso_98, macaluso_99}, who observed the
enhancement of  the Faraday  effect under  resonant conditions  in low
density material (atomic media).

Advances in laser technology and in laser spectroscopy, in conjunction
with intense studies on optical pumping processes \cite{happer_rmp_72,
  happer_book_10} prepared  a revival  of optical magnetometry  at the
beginning of  this millennium \cite{alexandrov_lp_96}. The  practicality and the  potential of
this research  in applications,  was induced by  further technological
advances, making available easy-to-use (reliable, low-cost, low-power,
small size, highly tunable and  stable) solid state laser sources, and
high-quality   atomic  samples   with  long   ground-state  relaxation
times. Many research  groups contributed to this new  phase of optical
magnetometry, and a panoramic view of that recent history can be found
in \cite{budker_natph_07}.

Optical  magnetometers are  developed  for a  variety of  applications
\cite{Savukov_intech_10}      including      fundamental      research
\cite{Swank_pra_18,  guarrera_prl_20, abel_pra_20},   characterization   of   magnetic
anomalies and  of their dynamics in  space-physics \cite{korth_jgr_16,
  pollinger_mst_18},               geology \cite{prouty_book_2013},              archaeology
\cite{fassbinder_encyarch_17, mathe_as_17},  material  science  e.g.  to
detect     diluted     magnetic      nano-     and     micro-particles
\cite{jaufenthaler_sens_20}      or     induced      eddy     currents
\cite{marmugi_apl_19,  jensen_prr_19, bevington_jap_19}   (with  potential   in  medical
applications, in  detection of  biomagnetism \cite{baranga_apl_06}  or in  building apparatuses
for nuclear  magnetic resonance (for  spectroscopy or imaging)  in the
ultra-low-field      \cite{savukov_prl_05,      biancalana_DH_jpcl_17,
  tayler_jmr_19,   biancalana_apl_19}   and  zero-to-ultra-low   field
regimes \cite{blanchard_jmr_20, xu_pnas_06}.

Among the most attractive characteristics  of OAMs is their robustness
and  the  possibility  of  pushing  to  extreme  levels  many  crucial
parameters  such  as  sensitivity, size,  minimal  power  consumption,
bandwidth, long-term operativity, etc.  The high sensitivity of atomic
magnetometers relies  on the  fact that  light near-resonant  with an
optical transition may create  long-lived magnetization in the atomic
ground  state  that  subsequently  evolves under  the  effect  of  the
magnetic field that is being measured.  This precession in turn modifies the
optical properties  of the atomic medium,  and can be  detected by
absorption  and/or  polarimetric  measurements performed  on  a  probe
radiation propagated through the medium itself.

Regardless of  the quantum or  the classical approach to  the problem,
the  evolution of  a  spin  precessing in  a  magnetic  field is  well
described by the Larmor equations, or by the Bloch equations when spin
relaxation  processes  play  important role.  This  magnetic-resonance
picture is  actually the framework in  which the spin dynamics  in OAM
can be described.

When  dealing  with  time-dependent   fields,  the  problem  is  often
considered in the  resonant regime, when a  field component oscillates
at the Larmor frequency (this is the case of radio-frequency
magnetometers).   Other studies,  concerning fast  oscillating fields,
consider the magnetic dressing phenomena, which address configurations
with a strong  field that oscillates at frequencies  largely above the
Larmor  one. At  the  other extreme,  the  quasi-stationary regime  is
analyzed (e.g. in the case  of light-modulated OAMs), which let detect
slow   and    extremely   weak   field    variations.

When considering a quasi-static TDF $\mathbf{b}$ much weaker than  the static field $\mathbf{B}_L$ to
which  it  is  superimposed,  the  scalar nature  of  OAM  makes  them
responsive  only  to  the  longitudinal TDF  component: the  Larmor frequency  $\omega_L$
depends  of the  field modulus  via the  gyromagnetic factor  $\gamma$
according to
\begin{equation}
    \omega_L=\gamma |\mathbf{B}|= \gamma \sqrt{(\mathbf{B}_L +  \mathbf{b})\cdot (\mathbf{B}_L +  \mathbf{b})}
    \approx 
    \gamma(B_L+\delta b_{\parallel}),
    \label{eq:firstorderstatic}
\end{equation}
to    a    first-order    Taylor    expansion    in    the    quantity
$b_{\parallel}/B_L$,  being   $b_{\parallel}$   the
component of $   \mathbf{b}$ along the direction of  $\mathbf{B}_L$.  The
high sensitivity of  OAM makes indeed these  detectors excellent tools
to detect very slow and extremly  weak TDFs.   The  weak-   and  low-frequency   field approximation   leading  to   Eq.~\ref{eq:firstorderstatic}  is   often
implicitly  assumed  \cite{wilson_prr_20, jensen_sr_18}:  the  dynamic  response  to stronger, generically  oriented and/or  non quasi-static TDF  remains
overlooked. Only few  works address operating conditions  with TDF not
necessarily  small,   with  fast  dynamics  and   generic  orientation
\cite{ingleby_pra_17,  wilson_prr_20}, and  this investigation  --with
emphasis to the system response-- is at the focus of the present work.

If we  consider a Bell and  Bloom magnetometer driven at  resonance, a
slight change of the magnetic field modulus will bring the device to a
near-resonant regime, with  a correspondent new (near-resonant) steady
state    characterized     by    smaller    and     dephased    atomic
magnetization.   However,   we  are   not   dealing   with  a   linear
time-invariant (LTI) system forced by  a signal initially resonant and
then out-of-resonance. The magnetometer should be rather regarded (see
also  the  appendix in  ref.  \cite{Savukov_mst_17})  as a  parametric
system forced by a stationary term. Despite the inherent similarity at
the steady  state, the transient response  of such a system  cannot be
studied in terms of a damped-forced (LTI)
device. 

Parametric  systems  are  used  in  a variety  of  sensors  e.g.   for
nanoscale  mass  and acceleration  detectors  \cite{villanueva_nl_11},
solid    state   gyroscopes    \cite{feng_ieee_04},   micropositioning
\cite{santhosh_pt_12}.  In  some  cases  instabilities  of  parametric
oscillators can  be used to sustain  the sensor \cite{surappa_apl_17},
which is viable technique when  the equation ruling the dynamics makes
it  possible to  provide the  system  with energy  from the  parameter
modulation. This is the case of many parametric oscillators, but it is
not a  general feature and  does not apply  to the particular  case of
precessing spins. The  dynamic response of a parametric  system to the
parameter(s) variation depends  strictly on its nature  and requires a
dedicated analysis of the equations that rule its dynamics.

This work provides an accurate  description of the dynamical behaviour
of   spins  precessing   in  a   TDF.   Similarly   to  the   case  of
Ref.\cite{ding_sens_18}, a  Bell and Bloom magnetometer  is considered
as  a test  of  the developed  model, under  conditions  in which  the
driving  field $\mathbf{B}$  changes  in time  with  TDF components  both
parallel and transverse  with respect to the static  (bias) term.  The
field is assumed to vary not necessarily slowly and by small but not necessarily vanishing amounts.  Differing  from Ref.\cite{ding_sens_18}, which reports
a numerical study,  in the Sec.~\ref{sec:model} we  develop an analytic
perturbative model to evaluate the response of such system to TDFs.

In   Sec.~\ref{sec:setup}   we  describe   an
experimental setup, where a weak synchronous optical pumping acting on
an atomic  vapour compensates the relaxation  phenomena, rendering the
evolution  of the  sample  magnetization adequately  described by  the
Larmor  precession  equation.   The  Sec.~\ref{sec:results}  reports  a
quantitative analysis  and experimental tests  of the main  outputs of
the model.  A  synthesis of the main findings is  finally drawn in the
Sec.~\ref{sec:conclusion}.

%% file: model.tex
\section{Model \label{sec:model}}

\newcommand{\MM}{\mbox{$\mathbf{M}$}}
\newcommand{\Mz}{\mbox{$\mathbf{M}^{(0)}$}}
\newcommand{\Mu}{\mbox{$\mathbf{M}^{(1)}$}}
\newcommand{\Md}{\mbox{$\mathbf{M}^{(2)}$}}
\newcommand{\re}{\mbox{$\mathrm{Re}$}}
\newcommand{\im}{\mbox{$\mathrm{Im}$}}

\begin{table*}[ht]
\begin{tabular}{|c|c|}
\toprule
symbol & description  \\
\hline
$\gamma$ & gyromagnetic factor \\
$\omega_L=\gamma |B_L|$ & Larmor angular frequency in the static field\\
$\omega$ & angular frequency of the modulating signal\\
$\Gamma$ & relaxation rate\\
$\mathbf {M}$ &  magnetization\\
$i=1,2,3$ & indexes corresponding to $x, y, z$ directions\\
$b_i=  \omega_i / \gamma$ & time dependent field components\\
$f(t) = \sum_n f_n \e^{ i n \omega t} $ & forcing term\\
$w_i$ & amplitudes of oscillating $b_i$\\
$a_k, \varphi_k$ & amplitude and phase of the polarimetric signal, at the $k$th order\\
$\Omega_{i,k}$ & Fourier angular frequencies of the $i$th component of $\textbf{b}$\\
\toprule
\end{tabular}
\caption{Symbols}
\label{tab:def:symbols}
\end{table*}

The starting point in  modeling of the spin response is the
Larmor equation for the magnetization vector
\begin{equation}
  \label{eq:Larmor:1}
  \dot{\mathbf{M}} = - \Gamma \mathbf{M} + \gamma \mathbf{B}(t) \times \mathbf{M} + \mathbf{f}(t)
\end{equation}
where  $\gamma$   is  the  gyromagnetic  factor   and  $\mathbf{f}(t)$
represents the action  of the circularly polarized  pump radiation and
it  is modeled  as a  forcing term  (a list  of symbol  definitions is
reported  in  Tab.\ref{tab:def:symbols}).  The  total  magnetic  field
$\mathbf{B}$ is  composed of a  large static one  $\mathbf{B}_L$ (bias
field)  perpendicular to  the pump  radiation wavevector  and a  small
TDF $\mathbf{b}(t)$.  A
simplified isotropic decay mechanism ($  - \Gamma \mathbf{M}$) is included.  This is
a     well     justified     approximation     as     discussed     in
\cite{biancalana_apb_16}. Fixing the  axis in such a way  that the $x$
axis  is along  the pump  and the  $y$ axis  is along  the bias  field
$\mathbf{B}_L$ one obtains the equations
\begin{subequations}
  \label{eq:Larmor:pm}
  \begin{align}
    \dot{M}_{+} & = - \Gamma M_{+}  - i ( \omega_L + \omega_2(t) ) M_{+}
    + ( i \omega_1(t) - \omega_3(t) ) M_2 + f(t) \\
    \dot{M}_{-} & = - \Gamma M_{-}  + i ( \omega_L + \omega_2(t) ) M_-
    - ( i \omega_1(t) + \omega_3(t) ) M_2 + f(t) \\
    \dot{M}_{2} & = - \Gamma M_{2} + 
    \frac{1}{2}( i \omega_1(t) + \omega_3(t) ) M_+ 
    -\frac{1}{2}( i \omega_1(t) - \omega_3(t) ) M_-
  \end{align}
\end{subequations}
where $M_{\pm} = M_x \pm i M_z$, $M_2 = M_y$, $\omega_L = \gamma B_L$ and
$\omega_i(t) = \gamma b_i(t)$. These equations must be solved in the
regime of $\omega_i(t) \ll \omega_L$, which can be rewritten in matrix
form as
\begin{equation}
  \label{eq:Larmor:matr}
  \begin{split}
    \dot{\mathbf{M}} & = - \Gamma \mathbf{M} + 
    \begin{pmatrix}
      - i \omega_L & 0 & 0 \\
      0 & i \omega_L & 0 \\
      0 & 0 & 0 
    \end{pmatrix} \mathbf{M} + \epsilon
    \begin{pmatrix}
      - i \omega_2 & 0 & i \omega_+ \\
      0 & i \omega_2 & -i\omega_-  \\
      i \omega_-/2 & -i \omega_+/2 & 0 
    \end{pmatrix} \mathbf{M} + 
    f(t)\begin{pmatrix}
      1\\
      1\\
      0
    \end{pmatrix} \\
    &= - \Gamma \mathbf{M} + 
    ( A_0 + \epsilon A_1(t) ) \mathbf{M} + f(t) \mathbf{u} 
  \end{split}
\end{equation}
here we have slightly redefined the magnetization 
as  $\mathbf{M} = (
M_+, M_-,  M_2)$ and introduced  $\omega_{\pm}(t) = \omega_1(t)  \pm i
\omega_3(t)$. The parameter $\epsilon$  is just
a bookkeeping device for the perturbation theory. 
In fact writing $\MM
= \Mz + \epsilon \Mu + \epsilon^2 \Md + \ldots $ we have
\begin{subequations}
\label{eq:Larmor:pert}  
\begin{align}
  \dot{ \mathbf{M} }^{(0)} &= - \Gamma \Mz + A_0 \Mz + \mathbf{u} f(t) \\
  \dot{ \mathbf{M} }^{(1)}   &= - \Gamma \Mu + A_0 \Mu + A_1(t) \Mz \\
  \dot{ \mathbf{M} }^{(2)}   &= - \Gamma \Md + A_0 \Md + A_1(t) \Mu 
\end{align}
\end{subequations}
The steady-state solution for \Mz can be  obtained noticing that
the function $f(t)$ is a real and periodic function (whose exact form is not
important as it is shown below) which represents the modulated pumping
\begin{equation}
  \label{eq:form:pumping}
  f(t) = \sum_n f_n \e^{ i n \omega t} 
\end{equation}
here $\omega$ ($\approx \omega_L$  in the experiment) is the frequency
modulation of the pumping laser. With standard methods one finds
\begin{equation}
  \label{eq:sol:M0}
  \Mz = \sum_n f_n \; \frac{1}{ \Gamma - A_0 + i n \omega} \mathbf{u} \; 
  \e^{i n \omega t}
\approx \begin{pmatrix}
\cfrac{f_{-1}}{\Gamma + i\; \delta} \e^{- i \omega t} \\
\cfrac{f_{1}}{\Gamma - i \;\delta} \e^{ + i \omega t} \\
0
\end{pmatrix}
\end{equation}
where  $\delta = \omega_L  - \omega$  is the  Larmor detuning  and the
second approximated form is obtained retaining only the resonant terms
from   the  first.   This   solution  represents   the  steady   state
magnetization due to the bias field $\mathbf{B}_L$ only. 

The quantity experimentally monitored is the phase of $x$ component of
the magnetization, that is if we 
write
\begin{equation}
  \label{eq:def:fase:exp}
  M_+(t) =  a(t)\e^{i \varphi(t)} \e^{-i \omega t} 
\end{equation}
then $\varphi(t)$ is observed  in the experiment. Accordingly with the
perturbation theory we can write
\begin{subequations}
  \label{eq:fase:pert}
  \begin{align}
    \varphi(t)  &=   \varphi_0  +  \epsilon   \varphi_1(t)  +  \epsilon^2
    \varphi_2(t) + \ldots \\
    a(t) &= a_0 + \epsilon a_1(t) + \epsilon^2 a_2(t) + \ldots
  \end{align}
\end{subequations}
Using \eqref{eq:sol:M0} it follows that $a_0 = |f_{-1}/(\Gamma + i
\delta)|$ and $\varphi_0 = \arg
(f_{-1}/(\Gamma + i\delta))  $ which is not interesting  because it is
just an offset in the experimental signal.

\subsection{First order solution}
Substituting  \eqref{eq:sol:M0} into \eqref{eq:Larmor:pert}  
 the steady-state first order solution is found:
\begin{equation}
  \label{eq:Larmor:pert:1ord}
  \Mu =  \e^{-(\Gamma -  A_0)t}\int_0^t \e^{(\Gamma -  A_0)t'} A_1(t')
  \Mz(t') \de t'
\end{equation}
and  due  to  the diagonal  form  of  the  $A_0$ matrix  the  detailed
expressions are
\begin{subequations}
  \label{eq:Larmor:pert:1ord:detail}
  \begin{align}
    M_+^{(1)}&= -i \frac{f_{-1}}{\Gamma + i \delta}
    \e^{-(\Gamma  + i \omega_L)t}\int_0^t  \e^{(\Gamma + i
      \omega_L - i \omega)t'} 
    \omega_2(t') \de t' \\
    M_-^{(1)} &= ( M_+^{(1)} ) ^* \\
    M_2^{(1)}&= \frac{i}{2} \frac{f_{-1}}{\Gamma + i \delta}
    \e^{-\Gamma t}\int_0^t  \e^{(\Gamma - i \omega)t'} 
    \omega_-(t') \de t' 
    - \frac{i}{2} \frac{f_{1}}{\Gamma - i \delta}
    \e^{-\Gamma t}\int_0^t  \e^{(\Gamma + i \omega)t'} 
    \omega_+(t') \de t' .
  \end{align}
\end{subequations}
From 
\begin{equation}
  \label{eq:M+:first:order}
  \begin{split}
    M_+ &\approx M_+^{(0)} + \epsilon M_+^{(1)} \\
    &=\frac{f_{-1}}{\Gamma + i \delta}\e^{-i \omega t} 
    \left[  1  -  i  \epsilon  \e^{-(\Gamma  +  i  \delta  )t}\int_0^t
      \e^{(\Gamma + i \delta)t'} \omega_2(t') \de t' 
    \right]
\end{split}
\end{equation}
and using the relations
\begin{subequations}
  \label{eq:def:arg:pert}
  \begin{align}
    \arg\left[ z_0 ( 1 + \epsilon u_1 + \epsilon^2 u_2) \right]
    & = \arg(z_0) + \im(u_1) \epsilon + \im( u_2 - u_1^2/2) \epsilon^2
    + \ldots \\
    \left| z_0 ( 1 + \epsilon u_1 + \epsilon^2 u_2) \right| 
    &  =  |z_0| \left[  1  + \re(u_1)  \epsilon  +  \left( \re(u_2)  +
        \frac{|u_1|^2 - \re(u_1)^2}{2} \right)\epsilon^2 + \ldots \right]
  \end{align}
\end{subequations}
one obtains 
\begin{subequations}
  \label{eq:phi1:def}
  \begin{align}
    u  &= -  i \e^{-(\Gamma  + i  \delta )t}\int_0^t  \e^{(\Gamma  + i
      \delta)t'} \omega_2(t') \de t'       \label{eq:phi1:u:def}\\  
    a_1(t) &= a_0 \re( u) \label{eq:phi1:a:def} \\
    \varphi_1(t) &= \im( u ) = 
    - \int_0^t \e^{-\Gamma ( t - t')} \cos( \delta (t - t') ) \; \omega_2(t') \de t' 
    \label{eq:phi1:phi:def}
  \end{align}
\end{subequations}

This result shows that the phase does not depend on the specific form of
the pumping signal. In fact, the $f_{\pm 1}$ coefficients  do not appear
in   Eq.\ref{eq:phi1:phi:def}.    The    same   equation   shows   that
$\omega_2(t)$ and $\varphi_1(t)$  can be thought as the  input and output
of a linear system with transfer function
\begin{equation} 
T(s) = -\frac{s + \Gamma}{ (s+\Gamma)^2 + \delta^2} =
  -\frac{1}{2}
  \left(
    \frac{1}{s + \Gamma - i \delta} + \frac{1}{s + \Gamma + i \delta}
  \right)
  \label{eq:firstorderresponse}
\end{equation}

Moreover  to  this  perturbative  order  the
spin response  is driven   only by the  component of  the small
magnetic  field  parallel  to  the  large  bias.  Finally,  the
expression    for    $\varphi_1$    agrees    with    that    reported
in \cite{biancalana_apb_16} in  the limit $\delta  \approx 0$, while
gives  more  precise results  in  agreement  with Ref.~\cite{zhang_applsc_20}.

For        instance,        for        a        sinusoidal        field
$\omega_2(t) = w \cos( \Omega t + \Phi )$ one obtains
\begin{equation}
  \label{eq:phi1:harmo}
  \varphi_1(t) = -\frac{w}{2} 
\left[
  \frac{1}{[\Gamma^2 +  ( \delta + \Omega)^2]^{1/2}} \cos(  \Omega t +
  \Phi - \psi_+) 
  + \frac{1}{[\Gamma^2 + ( \delta - \Omega)^2]^{1/2}} \cos( \Omega t +
  \Phi + \psi_-) 
\right] \qquad t \gg 1/\Gamma
\end{equation}
where $\psi_{\pm} = \arctan( (\delta \pm \Omega)/\Gamma )$.


\subsection{Second order solution}
The first order solution does not  depend on the TDF orthogonal to the
bias field  and one  has to  go one step  further in  the perturbative
expansion which is not difficult in principle, but the algebra becomes
quickly  a burden,  thus we  introduce a  simplifying hypothesis  very
close to  the experiment.   Let's assume that  the TDF  corresponds to
$\omega_i(t)$ in the form
\begin{equation}
  \label{eq:form:wi}
  \omega_i(t) = \sum_k c_{i,k} \e^{i \Omega_{i,k} t} \qquad \Omega_{i,
  -k} = - \Omega_{i,k} \;\;\; c_{i,-k} = c_{i,k}^*
\end{equation}
and  the  frequencies  satisfy   $\Omega_{i,k}  \ll  \omega$  (in  the
experiment $\omega / 2 \pi \sim 10 \, \mathrm{kHz}$ range while $\Omega_{i,k} / 2\pi
\sim 100 $ Hz range). With these assumptions we have for $t \gg 1/\Gamma $
\begin{subequations}
  \label{eq:explicit:1ord}
  \begin{align}
    \label{eq:exp:M1+}
    M_+^{(1)} & = -i \frac{f_{-1}}{\Gamma + i \delta} \e^{-i \omega t}
    \sum_k  \frac{c_{2,k}}{\Gamma +  i \delta  +  i\Omega_{2,k}} \e^{i
      \Omega_{2,k}t} = W(t) \e^{-i \omega t}\\
    M_2^{(1)} & = Z(t) \e^{-i \omega t } + Z^*(t) \e^ {i \omega t} \\
    Z(t) &= \frac{1}{2} \frac{f_{-1}}{\Gamma + i \delta} 
    \sum_k 
    \left[
      \frac{c_{3,k}}{\Gamma - i \omega + i \Omega_{3,k}} 
      \e^{i \Omega_{3,k} t}
      +
      \frac{ i c_{1,k}}{\Gamma - i \omega + i \Omega_{1,k}} 
      \e^{i \Omega_{1,k} t}
    \right]
  \end{align}
\end{subequations}
where  the  functions  $Z(t)$  and  $W(t)$ change  on  a  much  slower
timescale with respect to $\e^{\pm i \omega t}$. 

Writing $M_+^{(2)} = V(t) \e^{-i \omega t}$, the equation to solve reads as 
\begin{equation}
  \label{eq:2ord:M+}
  \begin{split}
    \dot{V} & = -( \Gamma + i \delta ) V - i \omega_2 W + ( i \omega_1 -
    \omega_3) (Z + Z^*\e^{2i\omega t}) \\
    &\approx -( \Gamma + i \delta ) V - i \omega_2 W + ( i \omega_1 -
    \omega_3) Z ,    
  \end{split}
\end{equation}
where the neglected term is a fast oscillating quantity.
The solution for $t \gg 1/\Gamma$ is 
\begin{equation}
  \label{eq:A:2ord}
  V = \frac{1}{2} \frac{f_{-1}}{\Gamma + i \delta} ( v_1 + v_2 + v_3
  + v_4  ) ,
\end{equation}
where
\begin{subequations}
\label{eq:ai:def}
  \begin{align}
    v_1 & = \sum _{k, k'}
    \frac{ -2 c_{2,k} c_{2,k'}}
    {(\Gamma + i\delta +i \Omega_{2,k})
      (\Gamma  +  i\delta +i  \Omega_{2,k'})}  \e^{i (\Omega_{2,k}  +
      \Omega_{2,k'}) t}  \label{eq:a1:def}
      \\
    v_2 & = \sum _{k, k'} \frac{ - c_{3,k} c_{3,k'}}
    {(\Gamma - i\omega +i \Omega_{3,k})
      (\Gamma + i\delta +i (\Omega_{3,k} + \Omega_{3,k'}) )} \e^{i (\Omega_{3,k} +
      \Omega_{3,k'}) t}   \label{eq:a2:def}
      \\
    v_3 & = \sum _{k, k'} \frac{ - c_{1,k} c_{1,k'}}
      {(\Gamma - i\omega +i \Omega_{1,k})
        (\Gamma + i\delta +i (\Omega_{1,k} + \Omega_{1,k'}) )} \e^{i (\Omega_{1,k} +
        \Omega_{1,k'}) t}\label{eq:a3:def}
        \\
    v_4 & = \sum _{k, k'} \frac{ (\Omega_{3,k'} - \Omega_{1,k}) c_{1,k} c_{3,k'}}
    {(\Gamma - i\omega +i \Omega_{1,k})(\Gamma - i\omega +i \Omega_{3,k'})
      (\Gamma + i\delta +i (\Omega_{1,k} + \Omega_{3,k'}) )} \e^{i (\Omega_{1,k} +
      \Omega_{3,k'}) t}  
    \label{eq:a4:def}
\end{align}
\end{subequations}
Using  again the Eq.\ref{eq:def:arg:pert},  after some  algebra we  find the
second order phase 
\begin{equation}
  \label{eq:phi2:def}
  \varphi_2(t)  = \im  \left( \frac{v_1}{4}  + \frac{v_2  + v_3  +  v_4 }{2} \right). 
\end{equation}

A close inspection shows that the $v_1$ term ``doubles'' and mixes the
frequencies present in $y$ component of  the small field (i.e. the TDF
component along the bias field). A similar behaviour is observed in the
terms  $v_2$ and  $v_3$: they  double and  mix the  frequencies of  TDF
components along the  $z$ and $x$ directions,  respectively.  Only the
$v_4$  term gives  rise  to frequencies  mixing  among orthogonal  TDF
components, and noticeably involves only  $x$ and $z$ terms: no mixing
occurs between transverse and longitudinal TDFs.

Notice also that, thanks to the
structure of  the denominators, in  the regime $\delta \approx  0$ the
$v_1$ term  is greater than the  others. Moreover the  $v_2$ and $v_3$
terms depend on the modulation frequency approximately as $1/\omega$ and the
cross-component mixing term is furtherly depressed, being $v_4 \sim 1/\omega^2$. 

Workable expressions can  be obtained in the case  of single frequency
TDF          applied      along     each     axis.     Substituting
$\omega_i(t) = w_i \cos( \Omega_i t  + \Phi_i) = ( w_i\e^{i \Phi_i} \;
\e^{i \Omega_i t } + w_i\e^{-i \Phi_i}  \; \e^{-i \Omega_i t })/2 $ in
Eq.\ref{eq:ai:def} one obtains, for instance:

\begin{equation}
  \label{eq:a1:simple}
  v_1 = -\frac{w_2^2}{2}\left( 
    \frac{\e^{-2  i (\Omega_2  t +  \Phi_2)}}{(\Gamma +  i \delta  - i
      \Omega_2)^2}
    +  \frac{\e^{2 i  (\Omega_2 t  + \Phi_2)}}{(\Gamma  + i  \delta +i
      \Omega_2)^2}
   + \frac{2}{(\Gamma + i \delta)^2 + \Omega_2^2}
  \right),
\end{equation}
which produces a peak at $\Omega = 2 \Omega_2$ in the square modulus  of the Fourier transform of $\varphi_2(t)$:
\begin{equation}
  \label{eq:a1:Fourier:peak:2harmy}
  |  \hat{\varphi}_2(   \Omega  =  2\Omega_2)   |^2  =  \frac{w_2^4}{4}   (\Gamma^2  +
  \Omega_2^2)\; 
  \frac{\delta^2}{(\Gamma^2   +  (\delta-\Omega_2)^2)^2   (\Gamma^2  +
    (\delta+\Omega_2)^2)^2}. 
\end{equation}
Similarly,
\begin{equation}
\label{eq:fourier:peaks:2harmxz}
| \hat{\varphi}_2( \Omega = 2\Omega_{1,3}) |^2  = \frac{w_{1,3}^4}{16} 
  \frac{ \Gamma^2 + 4 \Omega_{1,3}^2}{(\Gamma^2 + (\delta-2\Omega_{1,3})^2) (\Gamma^2 +
    (\delta+2\Omega_{1,3})^2)} \frac{1}{\omega^2} + O( \frac{1}{\omega^3})
\end{equation}
and, defining $\Omega_\pm=|\Omega_1\pm\Omega_3|$,
\begin{equation}
\label{eq:fourier:peaks:mixingxz}
| \hat{\varphi}_2 (  \Omega = \Omega_{\pm}) |^2   = \frac{w_3^2
    w_1^2}{16} 
  \frac{\Omega_{\mp}^2 (\Gamma^2 + \Omega_{\pm}^2 )}
  {( \Gamma^2 + ( \delta - \Omega_{\pm})^2)
    (   \Gamma^2    +(   \delta   +   \Omega_{\pm})^2)   }
  \frac{1}{\omega^4} + O( \frac{1}{\omega^5}).
\end{equation}
  It is convenient to consider the square-amplitude ratio among sum-frequency and difference-frequency terms
\begin{equation}
    R=\frac{\Omega_+^2 (\Omega_-^2+\Gamma^2) [\Gamma^2+(\delta-\Omega_+)^2][\Gamma^2+(\delta+\Omega_+)^2]}
           {\Omega_-^2 (\Omega_+^2+\Gamma^2) [\Gamma^2+(\delta-\Omega_-)^2][\Gamma^2+(\delta+\Omega_-)^2]},
    \label{eq:mixing:pmratio}
\end{equation}
which is more appropriate for experimental comparison. 

The              Eqs.\ref{eq:a1:Fourier:peak:2harmy},             \ref
{eq:fourier:peaks:mixingxz}  and   \ref  {eq:mixing:pmratio}  describe
quadratic  terms which  scale differently  with $\omega$.   The second
harmonic   response   to   TDF   component  along   the   bias   field
(Eq.\ref{eq:a1:Fourier:peak:2harmy}) vanishes at  $\delta=0$, but does
not  scale  with   $\omega$  and  hence  easily   exceeds  the  others
(Eq.\ref{eq:fourier:peaks:2harmxz})   out   of  this   condition.   In
contrast, being $\omega \gg  \Omega_{1,3}, \Gamma$ the cross-component
mixing   term  described   by the  Eq.\ref{eq:fourier:peaks:mixingxz}   is
extremely  weak   compared  to  both   the  $2\Omega_2$  and   to  the
$2\Omega_{1,3}$ terms.

\subsubsection{Not exactly orthogonal fields}
Suppose that  the TDF applied transveresely to the bias field are  not exactly  orthogonal to each other, but there  is some
tilting $\delta \theta = \theta_1+\theta_3$. In formula
\begin{align}
  \omega_1(t) &= w_1 \cos(\Omega_1 t + \Phi_1) \cos\theta_1 +
                w_3 \cos(\Omega_3 t + \Phi_3) \sin\theta_3 \\
  \omega_3(t) &= w_1 \cos(\Omega_1 t + \Phi_1) \sin\theta_1 +
                w_3 \cos(\Omega_3 t + \Phi_3) \cos\theta_3   
\end{align}
the previous result (Eq.\ref{eq:fourier:peaks:mixingxz}) generalizes to
\begin{equation}
\label{eq:fourier:peaks:mixingxz:notortho}
| \hat{\varphi}_2 (  \Omega = \Omega_{\pm}) |^2   = \frac{w_3^2
    w_1^2}{32} 
  \frac{\Omega_{\mp}^2 (\Gamma^2 + \Omega_{\pm}^2 )(1 + \cos(2\theta_1
    +2 \theta_3) )}
  {( \Gamma^2 + ( \delta - \Omega_{\pm})^2)
    (   \Gamma^2    +(   \delta   +   \Omega_{\pm})^2)   }
  \frac{1}{\omega^4} + O( \frac{1}{\omega^5})
\end{equation}
while another mixing term appears:
\begin{equation}
\label{eq:fourier:peaks:mixingxz:notortho:single:axis}
| \hat{\varphi}_2 (  \Omega = \Omega_{\pm}) |^2   = \frac{w_3^2
    w_1^2}{32} 
  \frac{ (\Gamma^2 + \Omega_{\pm}^2 )(1 - \cos(2\theta_1
    +2 \theta_3) )}
  {( \Gamma^2 + ( \delta - \Omega_{\pm})^2)
    (   \Gamma^2    +(   \delta   +   \Omega_{\pm})^2)   }
  \frac{1}{\omega^2} + O( \frac{1}{\omega^3}).
\end{equation}
The latter  scales with  a lower  power of $1/\omega$  so that  it may
become   easily    dominant,   also   for   small    $\delta   \theta$
values.     Interestingly,    taking into account  the mixing    terms
(Eq.\ref{eq:fourier:peaks:mixingxz:notortho:single:axis}),  the  ratio
$R$ does not depend on $\delta\theta$ and reads
\begin{equation}
  R=\frac{(\Omega_-^2+\Gamma^2)
    [\Gamma^2+(\delta-\Omega_+)^2][\Gamma^2+(\delta+\Omega_+)^2]}
  {(\Omega_+^2+\Gamma^2) [\Gamma^2+(\delta-\Omega_-)^2][\Gamma^2+(\delta+\Omega_-)^2]}
  \label{eq:mixing:pmratio:notortho}
\end{equation}

%% file: setup.tex
\section{Experimental setup}
\label{sec:setup}

\begin{figure}[ht]
   \centering
    \includegraphics [angle=0, width= 75 mm] {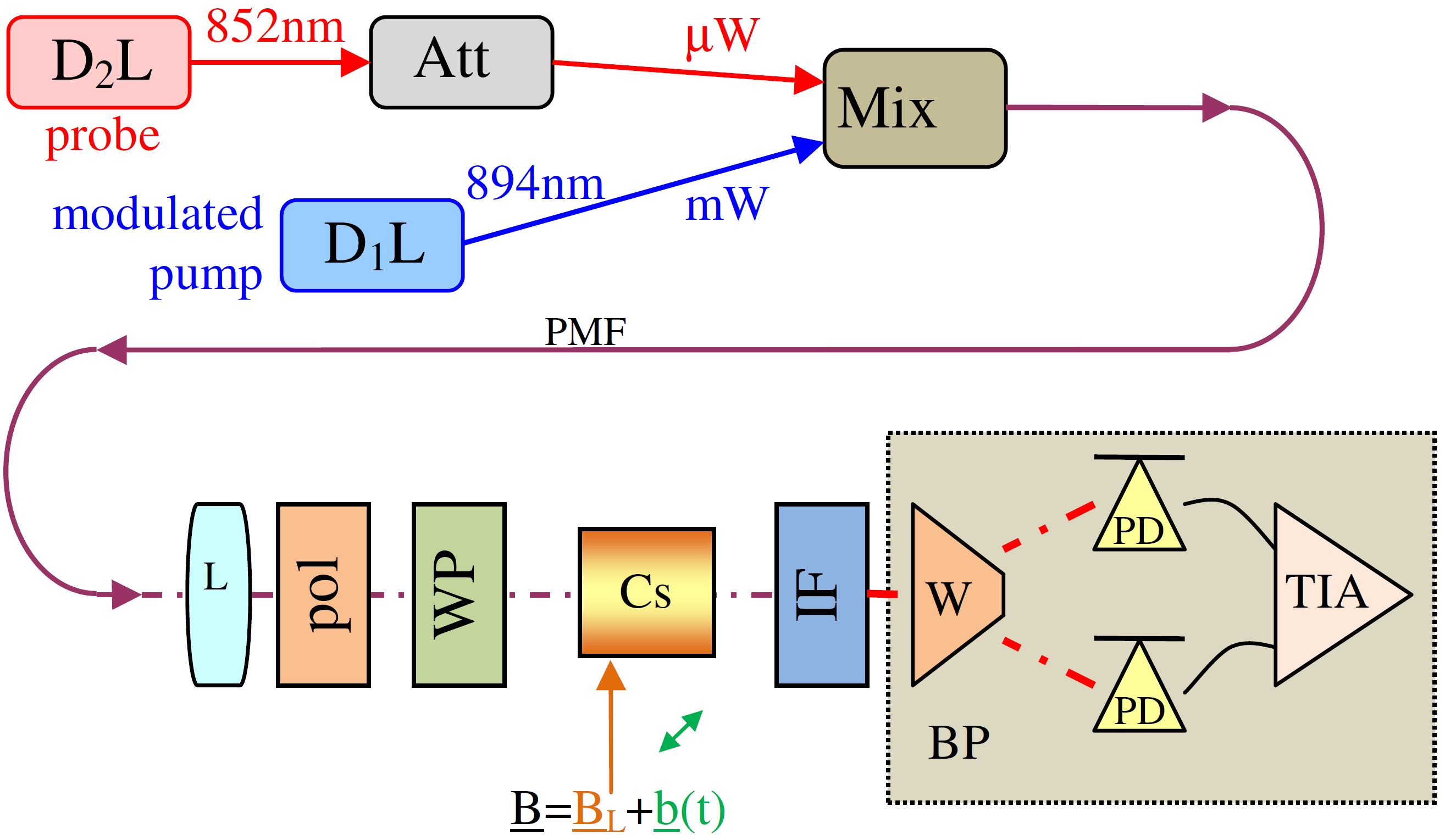}
    \caption{Schematics of the magnetometer. Two laser sources ($D_1L$
      and  $D_2L$) produce  pump  and probe  radiations, at  different
      wavelengths, which are mixed (Mix) and coupled to a polarization
      maintaining  fiber  PMF.   The   $D_1L$  wavelength  is  broadly
      modulated at  $\omega$ and maintained at  milliwatt level, while
      $D_2L$  is unmodulated  and attenuated  (Att) down  to microwatt
      level. At the PMF output, a  lens (L) collimates the radiation on a
      10~mm  diameter  beam,  whose  polarization is  reinforced  by  a
      polarizer  (pol)   and  modified  by  a   multi-order  waveplate
      (WP). The latter renders the pump radiation circularly polarized
      while leaving the probe  radiation linearly polarized. After the
      interaction  with  the atomic  sample  (Cs)  the pump  light  is
      blocked  by   an  interference   filter  (IF),  and   the  probe
      polarization is  analyzed by a balanced  polarimeternmade  by  a  Wollaston  prism (W)  and  a  pair  of  silicon
      photodetectors (PD). The photocurrent imbalance is  converted to a
      voltage  signal  by  a  transimpedance amplifier  (TIA),  to  be
      acquired  by a  16~bit  500~kS/s  card for  subsequent numerical
      elaboration. The  Cs cell is in  a magnetic field composed  by a
      large static  term $\mathbf{B}_L$  transverse to the  laser beam
      and  by  a weaker,  generically  oriented,  time dependent  term
      $\mathbf{b}$.}
  \label{fig:setup}
\end{figure}

The  dynamics  of atomic spins precessing
in  a  TDF  is  experiemtally studied by means of one  channel  of  the  multi-channel  Bell  and  Bloom
magnetometer   described  in   Ref.  \cite{biancalana_apb_16}.   Basic
information of the device is here summarized in Fig.~\ref{fig:setup}.

Beside a self-oscillating mode \cite{higbie_rsi_06, belfi_josa_09} not
relevant for  the scopes of  this work,  the magnetometer can  be used
under \textit{scan} or  \textit{forced} modes. In the  first case, the angular
 frequency $\omega$ of the pump laser modulation is scanned  around the atomic magnetic resonance,
to characterize  center, amplitude, width  and shape of  the resonance
itself, while in the \textit{forced} mode,  the modulation is set at a
(near) resonant  frequency, and TDF  is detected via the  phase shifts
induced in the polarimetric signal.
The  measurements  described  in  this   paper  are  obtained  in  the
\textit{forced} mode, after having run the system in the \textit{scan}
mode to determine the resonant modulation frequency and the resonance width $\Gamma$.

The   sensor  is   operated  in a magnetically  unshielded environment, where the Earth field is  partially   compensated
 by means of three mutually orthogonal Helmholtz coils. 
Additional coils  complete the setup to  apply variously oriented TDF  that   are normally used   for  the  manipulation  of   atomic  spins
\cite{biancalana_pra_12, biancalana_apl_19, biancalana_prappl_19}.

A solenoid surrounding the atomic cells is used to apply a homogeneous
TDF   along   the   propagation   direction   of   the   laser   beams
($x$).  Helmholtz pairs  or far  located dipoles  are used  to produce TDF
components along the static field ($y$) and in the other perpendicular
direction ($z$).  These TDF sources  can be supplied by  RF generators
configured   as  voltage   generators  with   series  resistors.   The
frequencies  of  the  applied  signals  are low  enough  to  make  the
inductive nature of the loads  negligible. A simplified calibration is
performed  for each  field  source under  magnetostatic conditions  to
determine the  voltage-to-field conversion coefficients.  The complete
control  of static  and  time-dependent field  components enables  the
analysis of the system response at the focus of the present work.

%% file: discussion.tex
\section{Discussion}
\label{sec:results}

\subsection{First-order approximation}

The  perturbative approach  presented in  Sec.\ref{sec:model} confirms
that the response to small and quasi-static  ($\Omega \ll \Gamma$) TDF is
consistent       with        the       approximation       anticipated
in Eq.~\ref{eq:firstorderstatic} and extends the analysis to the case of
TDF lower than $\omega_L/\gamma$, but with a dynamics non-necessarily slower than $\Gamma$ (i.e.
quasi-static).  In other words, it is only required that the  TDF is much weaker than the bias field. When this condition is strictly fulfilled (i.e. the TDF is extremely weak), the first perturbation order is sufficient to describe the system behaviour, and it shows that the system is  still responsive  to  the longitudinal  TDF
component only. Additionally, in the first-order approximation the dynamics of the
precessing     spins    can     be    described     in    terms     of
Eq.~\ref{eq:firstorderresponse},  referring  to the  notation  commonly used  for linear  systems, despite the parametric nature of the problem. 

At  the  steady state  ($s=i \Omega$),  the
mentioned response function reads:
\begin{equation} 
T( \Omega) = 
  -\frac{1}{2}
  \left(
    \frac{1}{ \Gamma + i (\Omega-\delta)} + \frac{1}{ \Gamma + i (\Omega+\delta)}
  \right),
  \label{eq:firstordersteadystate}
\end{equation}
which for  $\delta=0$,  corresponds  to the response of a
1-st order  Butterworth low pass  filter (the  same as an  RC circuit)
\cite{colombo_oe_16,  zhang_ieee_18},  while  for non-zero  values  of
$\delta$ (particularly for $|\delta| \gg \Gamma$) the  system responds
as a bandpass filter approximately centred at $\delta$.

\begin{figure}[ht]
   \centering
    \includegraphics [angle=0, width= 75 mm] {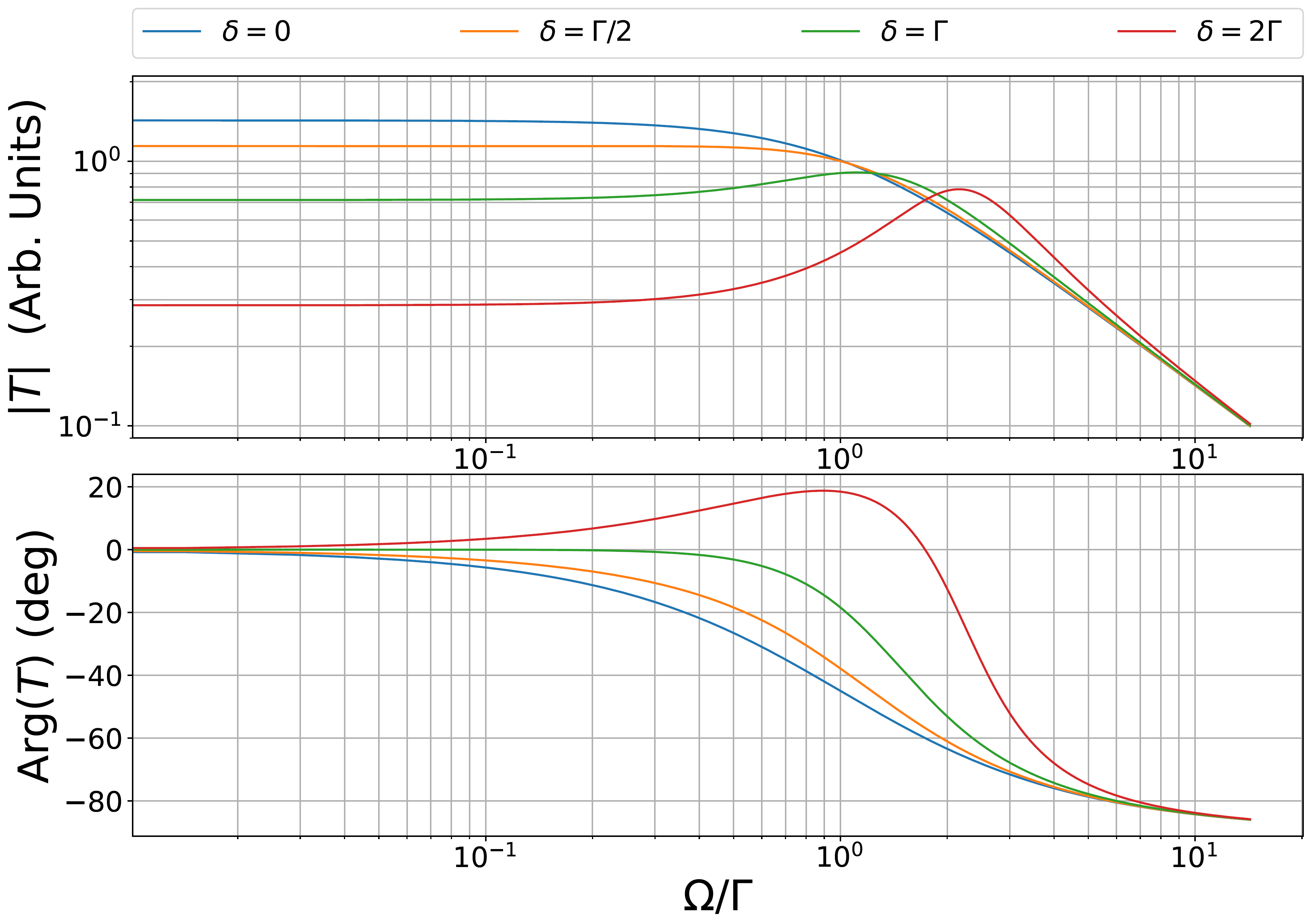}
    \caption{Theoretical Bode plot of the first order response.         }
  \label{fig.TiDiEsseBodeTheo}
\end{figure}

An  interesting feature  is obtained  under an  intermediate condition
($\delta \approx \Gamma/2$), which produces  a nearly flat response up
to a  cut-off frequency set  by $\Gamma/2$ itself. Such  extended flat
bandwidth  is  obtained at  expenses  of  a  slight reduction  of  the
response amplitude, if compared to the $\delta=0$ case. Similarly, the
condition   $\delta  =   \Gamma$   leads  to   a  maximally   extended
constant-phase  response.   Fig.~\ref{fig.TiDiEsseBodeTheo}  summarizes
these aspects showing the theoretical  Bode plot corressponding to the
Eq.~\ref{eq:firstordersteadystate},   for  four   relevant  values   of
$\delta$.
Both  the  extended flat  gain  ($\delta  \approx \Gamma/2$)  and  the
extended  flat phase  ($\delta \approx  \Gamma$) conditions  can be  of
interest  in magnetometric  applications, e.g.  in the detection of  magnetic
signals  with spectral  components which  range from  zero to  (about)
$\Gamma$ or when  the magnetometric signal is used to feed a closed-loop system
for active field stabilization \cite{biancalana_prapplml_19, zhang_sens_20}.

\begin{figure}[ht!]
   \centering
     \includegraphics[width=75 mm] {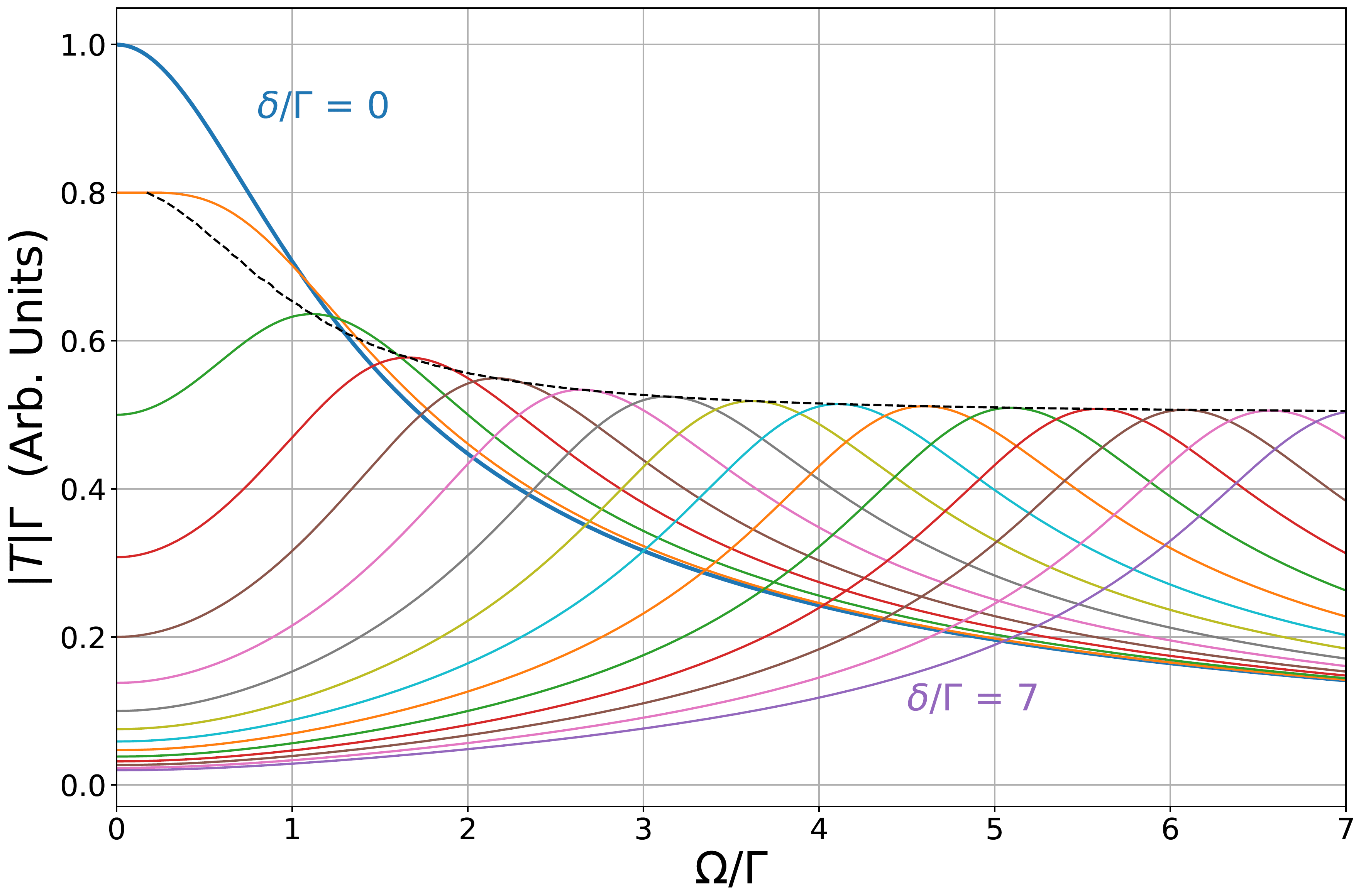}
     \caption{ 
         Amplitude of  the first  order response,
         for different  values ($\delta/\Gamma$ increases in  steps of
         1/2) of the detuning $\delta$. For large
         $\delta$     the     response     has    a     maximum     at
         $  \Omega_M  \approx  \delta$  and  noticeably  these  maxima
         converge to the asymptotic value $1/(2\Gamma)$ differing from
         the case $\delta=0$ for which a low-pass ($|T|\rightarrow 0$)
         response is found.
         Notice that  the maxima  are higher  than the  low-pass curve
         (blue thick line).
         }
  \label{fig:TM}
\end{figure}
Finally, the large $\delta$  regime can be of interest
in applications
that  need  an enhancement  of  the   response  to  TDF  oscillating  at
frequencies around $\delta$.

As said, for small values of  $\delta$ the response modulus $|T|$ is a
monotonically   decreasing   function    of   $\Omega$.    Then,   for
$\delta/\Gamma > \left(\sqrt{5}-2\right)^{1/2} \approx 1/2$, $|T|$ has
a                  maximum                 located                  at
$\Omega_M=\left                    (\sqrt{\delta^4+4\delta^2\Gamma^2}-
  \Gamma^2\right)^{1/2}$, which  turns to $\Omega_M  \approx |\delta|$
if $\delta \gg \Gamma$, i.e. when the band-pass regime occurs.
The maximum value of $|T|$
(let it be $T_M$)
is 
\begin{equation}
\label{eq:maxima:envelope}
T_M^2
=\frac{1}{8 \Gamma^2}\left(1+\sqrt{1+\frac{4 \Gamma^2}{\delta^2}}\right),
\end{equation}
which for  large values of  $\delta$ (and hence for  large $\Omega_M$)
does  not  decrease to  zero  (as  in  the  case of  $\delta=0$),  but
approaches  an asymptotic  value  $1/(2\Gamma)$.  Notice  that in  the
band-pass  regime the  response  obtained  for $\Omega\approx  \delta$
exceeds the response for the same $\Omega$ in the low-pass regime. 
This  behaviour is  summarized  in Fig.~\ref{fig:TM},  where a set  of
curves $T(\Omega)$  corresponding to  different detuning  $\delta$ are
plotted   together   with   the   enveloping   curve   described   by
Eq.~\ref{eq:maxima:envelope}.
The possibility of operating in a band-pass regime, with maximal system response centered at a selectable frquency has an evident  relevance in magnetometric applications where narrowband signals must be detected. As an  example, when an atomic
magnetometer     is    used     to    detect     nuclear    precession
\cite{biancalana_zulfJcoupling_jmr_16, biancalana_apl_19}, the nuclear
signal is narrowband in nature,  with a predictable frequency. In such
case,  an appropriate  selection  of $\delta$  let  enhance the system response to the signal under investigation.  

\begin{figure}[ht]
   \centering
    \includegraphics [angle=0, width= 75 mm] {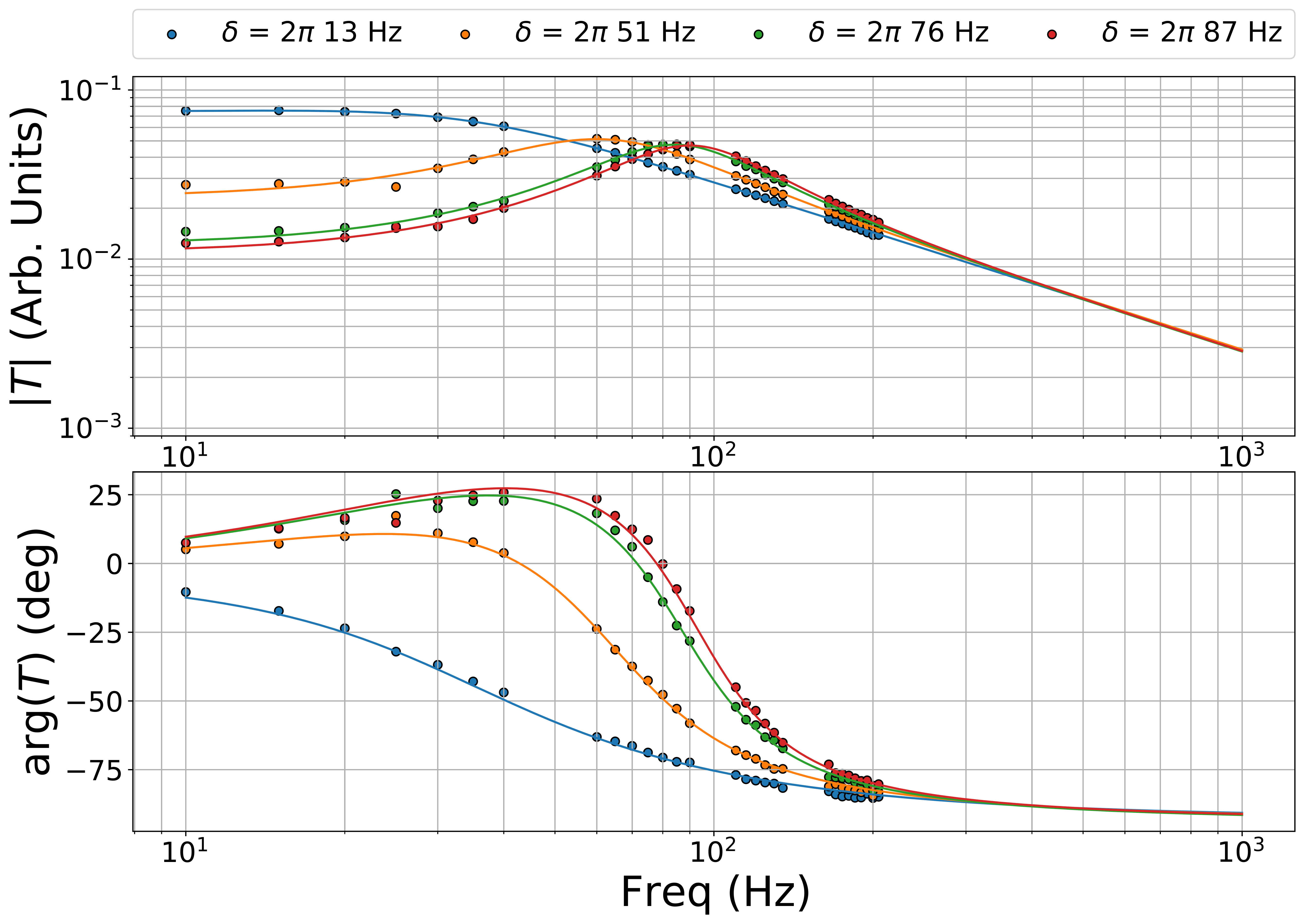}
    \caption{Bode plots of the first order response, theoretical (solid
      lines)  and  experimental  (points)  results,  as  obtained  for
      different values of $\delta$. Experimental points at 50~Hz, 100~Hz
      and  150~Hz   have  been  skipped,  because   affected  by  mains
      disturbance. The  response passes progressively from  a low-pass
      to  a  band-pass  behaviour, with  an  interesting  intermediate
      condition  ($\delta \approx  \Gamma  /2$), where  a nearly  flat
      response extends from  zero up to $\approx \delta$:  this is the
      case of the plot with $\delta=2 \pi\cdot 13$~Hz).
    }
  \label{fig.TiDiEsseBode}
\end{figure}

These predictions are validated with the apparatus described in 
Sec.\ref{sec:setup}.
The    Fig.~\ref{fig.TiDiEsseBode}    shows several
experimental   data sets  (amplitude and phases)  obtained with
weak TDF  applied along  the static  field direction,  with frequencies
ranging in a broad interval. Theoretical curves obtained from the model
(solid lines) are drawn with  the corresponding colours. Both amplitude
and phase responses  are considered for different  values of $\delta$,
and                          particularly                          for
$\delta \ll  \Gamma, \delta \sim  \Gamma, \delta > \Gamma,  \delta \gg
\Gamma $, and excellent agreement is found for all the regimes.

\subsection{Second-order approximation}

When  more intense  TDFs are  applied, the  second order  perturbation
terms  start  playing  a  role.  We  have  experimentally  tested  the
developed  model  by  applying  sinusoidal   TDFs  along  one  or  two
directions, which are nominally parallel  or perpendicular to the bias
field.  When   two  oscillating  components  are   applied,  different
frequencies  are  selected,  in  order  to  make  their  contributions
spectrally distinguishable.

At the second-order approximation, the model shows that the system response contains
quadratic  terms  of  both  the longitudinal  ($i=2$)  and  transverse
($i=1,3$) TDF  components. 
The simultaneous application of TDF with a single Fourier
component along different  directions $i$ ($i=1, 2,  3$, the direction
$2$ being that  of the bias field) with  different angular frequencies
$\Omega_{i}$, causes  the presence of  terms at $2 \Omega_{i}$  and at
$|\Omega_{1} \pm \Omega_{3}|$, while no mixing between transverse
and longitudinal TDF is expected.

It is worth noting that slight coil misalignments may lead to ``spurious'' mixing response
as   discussed   at  the   end   of   Sec.\ref{sec:model}  (Eqs.
\ref{eq:fourier:peaks:mixingxz:notortho}-\ref{eq:mixing:pmratio:notortho}),
and in applications this mixing can be used as a monitor tool, to
refine the coil orthogonality.

\begin{figure}[ht]
  \centering
  \subfloat[\label{fig:2Harm_xz_a}]{ \includegraphics [angle=0, width= 75 mm] {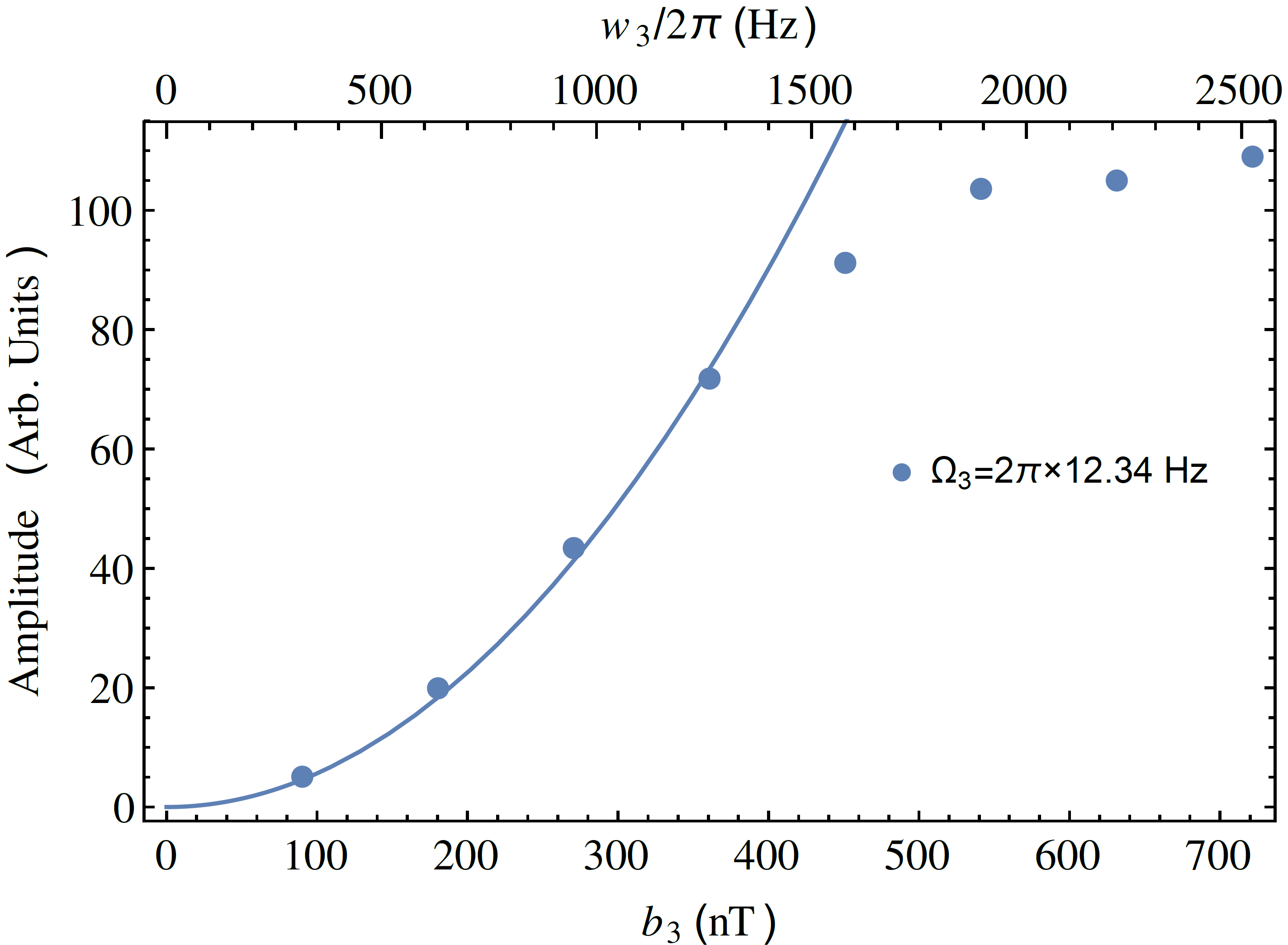}}
  \subfloat[\label{fig:2Harm_xz_b}]{ \includegraphics [angle=0, width= 75 mm] {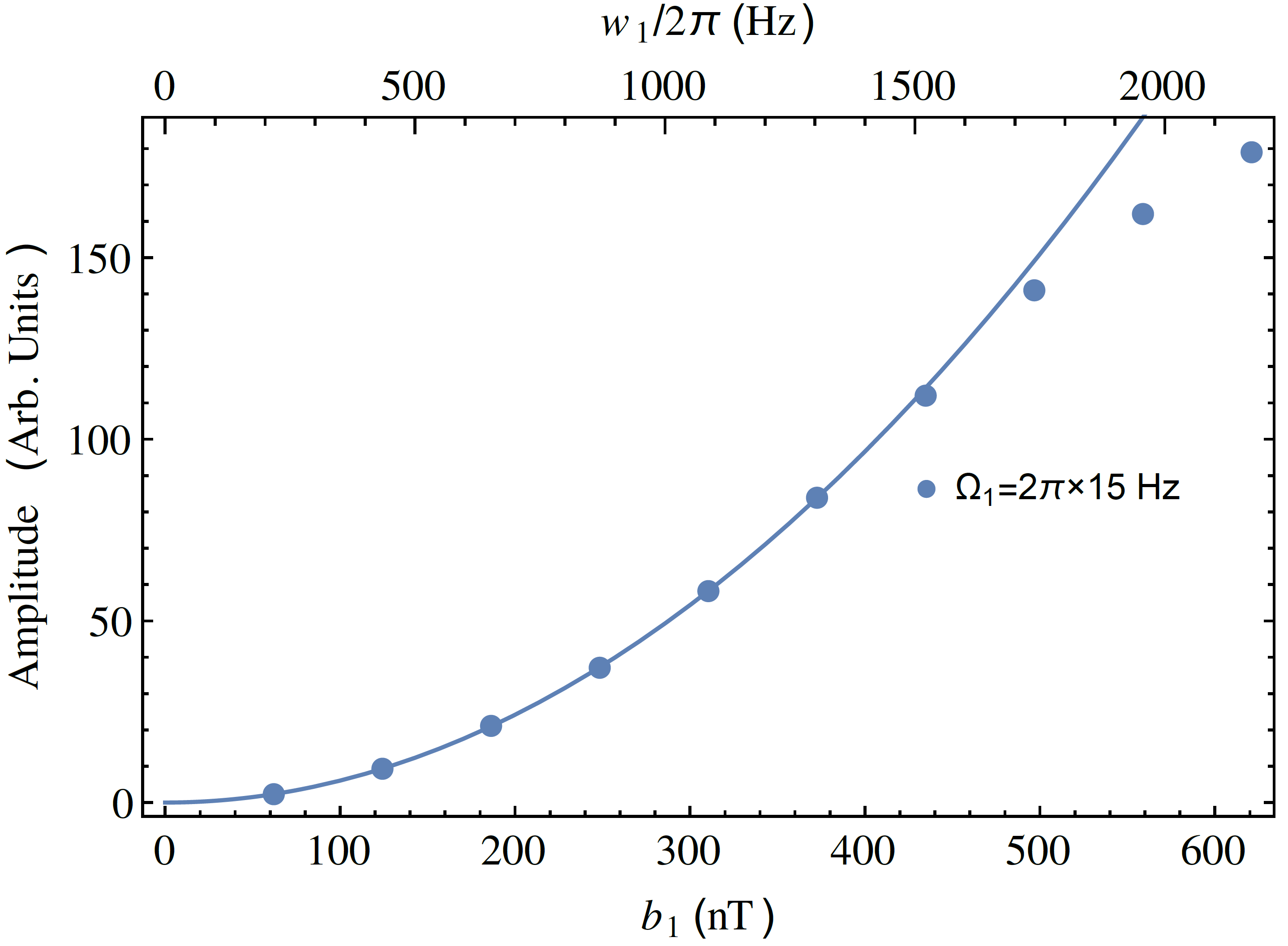}}
  \caption{ Quadratic  response to the transverse  fields. These plots
    report   the  second   harmonic  amplitude   registered  when   an
    oscillating transverse field is applied. In the case (a) a 12.34~Hz
    field       is        applied       along        $z$       ($b_3$,
    see Eq.~\ref{eq:fourier:peaks:2harmxz}a),  and in  the case  (b) a
    15~Hz    field    is    applied     along    $x$    ($b_1$,    see
    Eq.~\ref{eq:fourier:peaks:2harmxz}b).  A  saturation effect appears
    above  300-400~$\mu$T, while  the experimental  points recorded  at
    lower field  intensity are perfectly  fitted by a  parabolic curve
    (solid line).   Only the detuning  $\delta$ is varied in  the best
    fit  procedure, while  $\Gamma$  is set  a $2  \pi$30~Hz. The arbitrary units used for the vertical scales are the same as for the Figs.\ref{fig:2Harm_y} and  \ref{fig:mixingxz}.
      }
  \label{fig:2Harm_xz}
\end{figure}

On  the basis  of  Eq.~\ref{eq:fourier:peaks:2harmxz}, second  harmonic
peaks at  $2\Omega_1$ and  $2\Omega_3$ are  expected, when  one single
frequency  TDF  is  applied  along  the  $x$  or  $z$  direction.  
The Fig.~\ref{fig:2Harm_xz}  shows  experimental points  and  corresponding
fitting curves obtained  when a single frequency TDF  is applied along
$x$  or  $z$.  The  fitting    are
calculated  for an  assigned $\Gamma=2  \pi \cdot 30$~Hz, with  only one  free
parameter $\delta $ that is determined  to be $\approx\Gamma$, consistently with the experimental conditions. 

When the TDF is applied in the same direction as the bias field, the system second-order response is
described  by  the  Eq.~\ref{eq:a1:Fourier:peak:2harmy}:  the  expected amplitude of second-harmonics peak 
  may  noticeably   vary  as  a  function   of  $\delta$.  In
particular,  it is expected to vanish for  $\delta=0$: a feature that may find application  in stabilization systems  aimed to maintain  the  static  field  correctly oriented  and  under  resonant condition ($\omega_L \approx \omega$).

\begin{figure}[ht]
   \centering
    \includegraphics [angle=0, width= 75 mm] {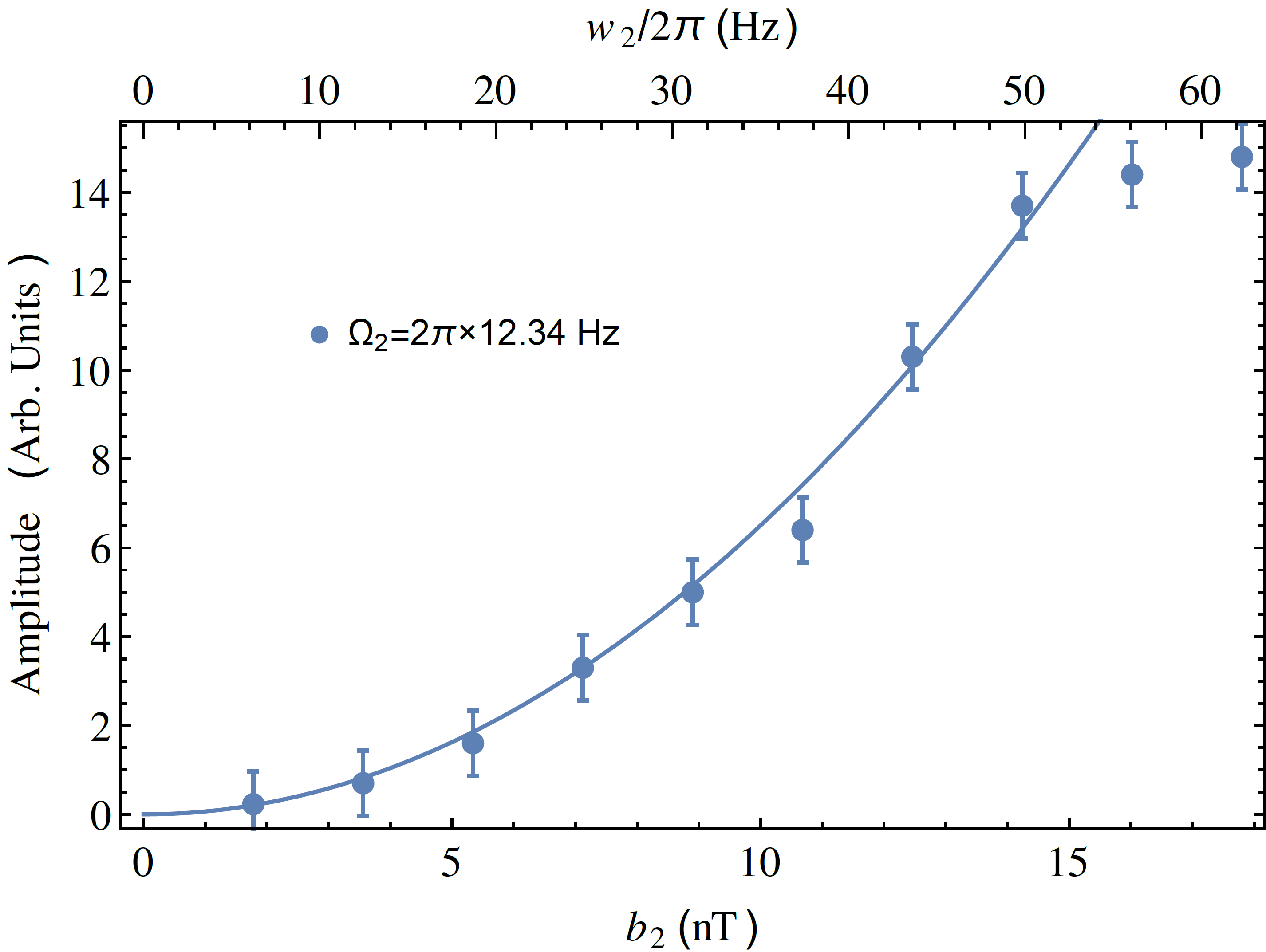}
    \caption{  In accordance  with Eq.~\ref{eq:a1:Fourier:peak:2harmy},
      provided  that a  detuning  $\delta$ exists,  a second  harmonic
      response with quadratically dependent  amplitude is expected, in
      response   to  longitudinal   TDF.  This   figure  reports   the
      $2\Omega_2$  peak registered  when a  12.34~Hz oscillating  $b_2$
      field is  applied. The relaxation rate is set at $\Gamma=2\pi\cdot30$~rad/s, and good fitting is found   for $\delta=$50~Hz,  which is
       consistent with the experimental conditions. In this case saturation is observed above 15~nT}
  \label{fig:2Harm_y}
\end{figure}

Fig.~\ref{fig:2Harm_y}  shows experimental  points and  a corresponding
fitting  parabola obtained when a TDF is applied along the bias field with different amplitudes. The fitting curve  lets  estimate
$\delta  \approx  50$~rad/s,  a value  consistent  with  the experimental conditions.

\begin{figure}[ht]
   \centering
     \includegraphics [angle=0, width= 75 mm] {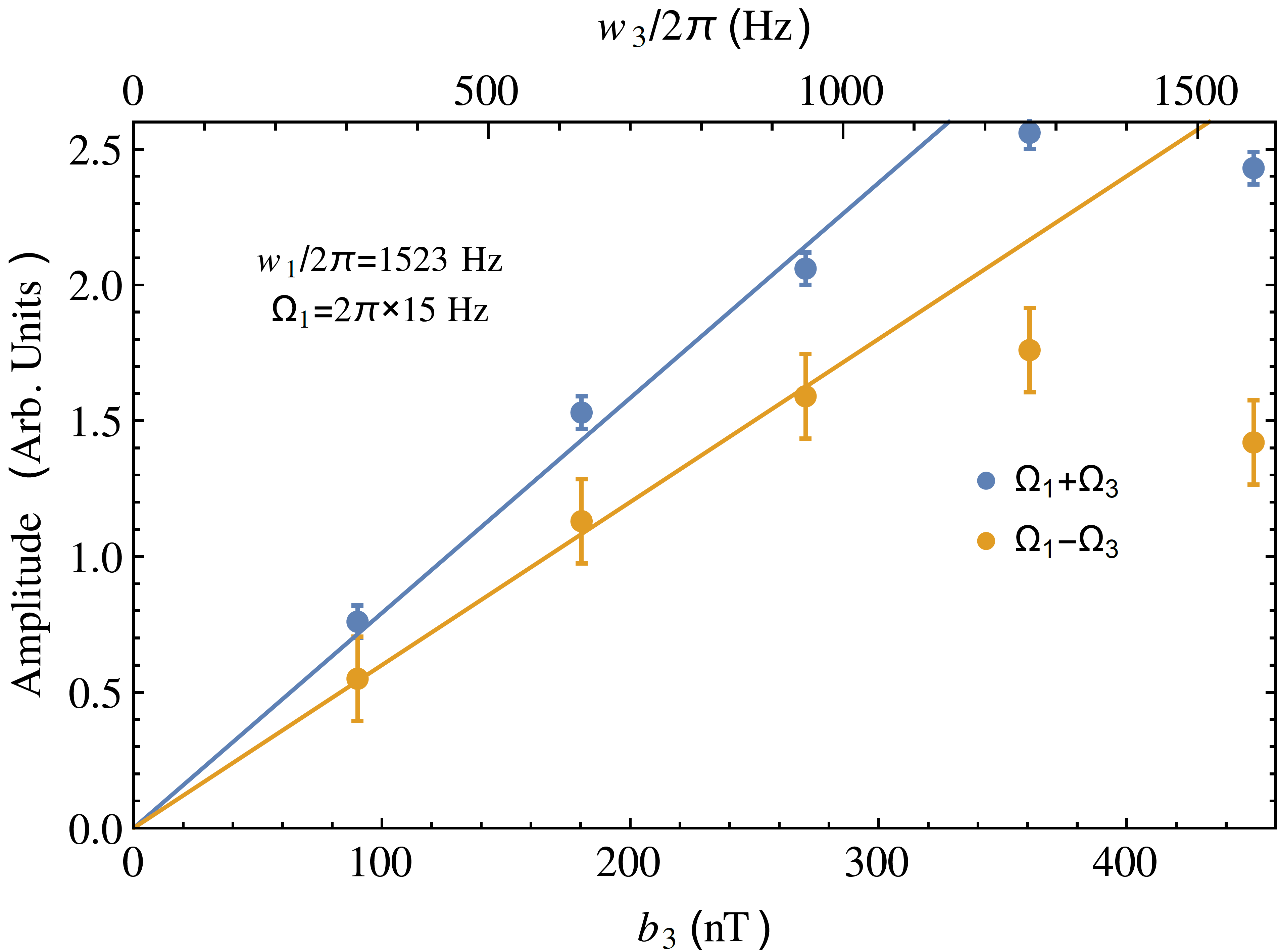}
    \includegraphics [angle=0, width= 76 mm] {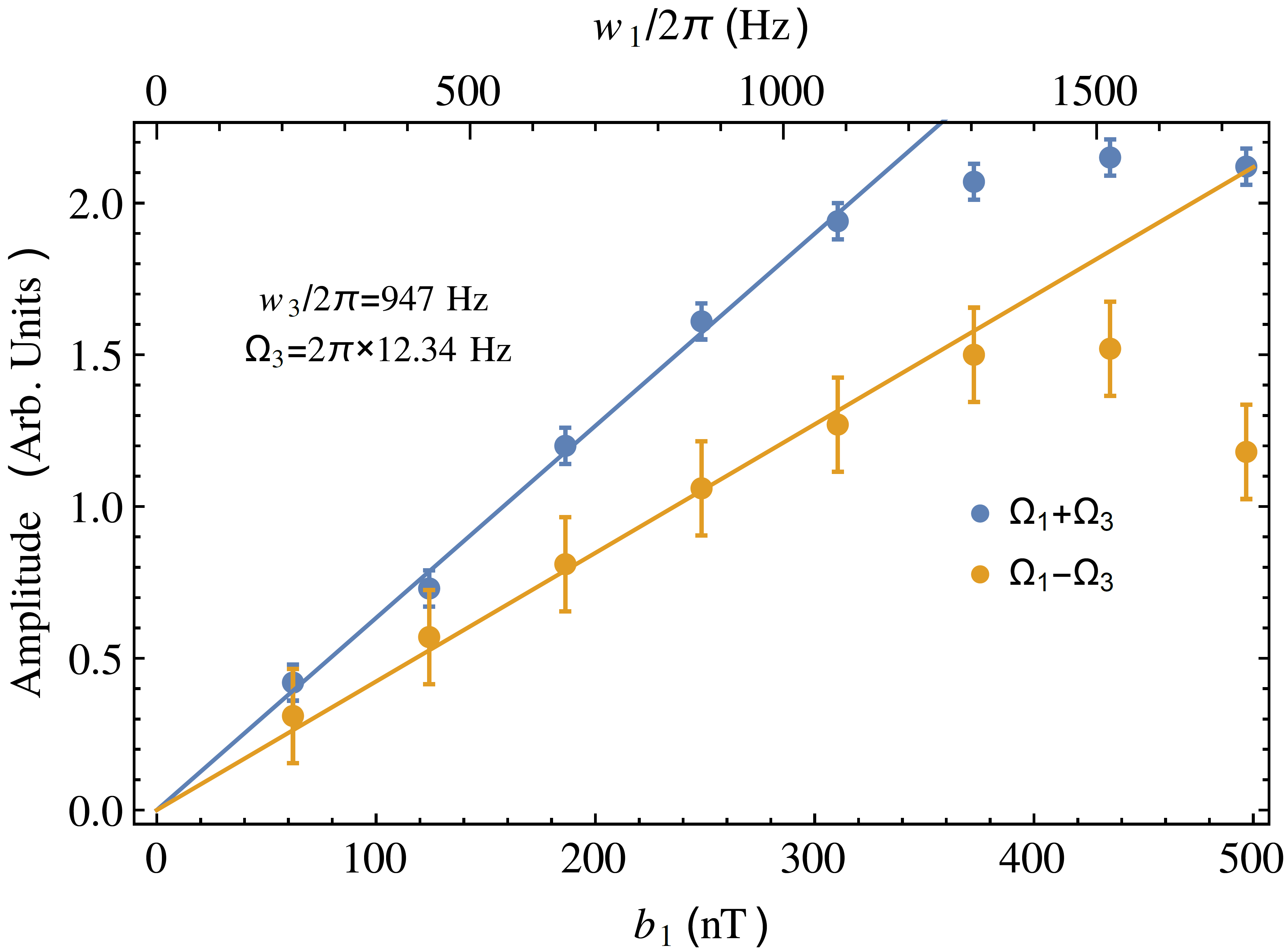}
    \caption{Linear dependence of mixing terms  on the $b_1$ and $b_3$
      amplitude. The two plots are obtained when oscillating fields at
      15~Hz and 12.34~Hz are  simultaneously applied along the $x$ and
      $z$ directions, respectively. In the  upper plot $b_1$ is set at
      435~nT and $b_3$ is varied, while  in the lower plot $b_3$ is set
      at 270.6~nT and $b_1$ is varied.  Peaks at the sum and difference
      frequencies  are recorded.  Both the  absolute amplitude  of the
      peaks and their ratios indicate that these mixed frequency peaks
      do  not  originate  from  the second  order  term  described  in
      Eq.~\ref{eq:fourier:peaks:mixingxz},  but  are rather  caused  by
      imperfect  coil  orthogonality.  The  ratio  between  sum-  and
      difference-frequency peaks  is indeed perfectly  consistent with
      the   square    root   of    the   ratio   $R$    expressed   in
      Eq.~\ref{eq:mixing:pmratio:notortho}. Similarly to the cases shown
      in  Fig.~\ref{fig:2Harm_xz},  a  saturation effect  occurs  above
      about 300~nT.  }
  \label{fig:mixingxz}
\end{figure}

The Eqs.~\ref{eq:fourier:peaks:mixingxz}                                and
\ref{eq:fourier:peaks:mixingxz:notortho:single:axis} describe a linear
dependence of the sum- and difference-frequency peaks as a function of
either the $x$ or $z$ TDF component. This behaviour is consistent with
the data shown in Fig.~\ref{fig:mixingxz},  up to TDF amplitudes around
300~nT.  Those data  are recorded when  applying two sinusoidal TDFs
along the $x$ and $z$ directions, at 15~Hz and 12.34~Hz, respectively.

The  linear slopes  of the  fitting curves  largely exceed  the amount
expected  from   Eq.~\ref{eq:fourier:peaks:mixingxz},  suggesting  that
orthogonality                                            imperfections
(Eq.~\ref{eq:fourier:peaks:mixingxz:notortho:single:axis})    play    a
dominant  role  in   this  case.  The  ratio  between   the  sum-  and
difference-frequency  slopes  is  approximately  1.4,  which  is  well
consistent       with       the      values       estimated       from
Eq.~\ref{eq:mixing:pmratio:notortho}   for  $\delta   \approx  0$   and
$\Gamma \approx 2\pi \cdot 30$~Hz.

%% file: conclusion.tex
\section{Conclusion}
\label{sec:conclusion}

We have studied the dynamic response of a parametric system constituted by a light-modulated atomic magnetometer. Such kind of instrumentation is  commonly used to detect extremely weak and slowly varying fields. Under these conditions, these devices are often analyzed with an implicit assumption, leading to treat them in analogy with linear time-invariant systems. We have developed a model that provides a more detailed analysis, and particularly let determine the response to TDFs that can range in a wider frequency interval and may have larger amplitudes. 

In a first order approximation, which is valid for tiny TDF, we find that the system responds only to the TDF component along the bias field, and that it acts as a low-pass or a band-pass filter, in dependence of the frequency mismatch between the Larmor precession and the pump laser modulation signal. Conditions to achieve maximally flat spectral response, or frequency independent dephasing are identified.

Extending the model to the second order approximation permits to analyze the system behaviour when larger TDFs are applied. Our findings show that the system responds quadratically to both longitudinal and transverse TDF components. The model provides quantitative evaluations of the diverse coefficients that describe amplitudes and phases of those nonlinear terms. In addition, frequency mixing may emerge between harmonic TDF components applied along the two transverse directions. The mixing occurs at different levels, depending on the relative orientation of the oscillating fields. In application (particularly when high resolution magnetometry is performed in the presence of narrow-band disturbances), this analysis will help to identify the origin of spurious spectral peaks that constitute artefacts in the detected signals.